\documentclass[12pt,preprint]{aastex}
\usepackage{psfig}

\def\spose#1{\hbox to 0pt{#1\hss}}

\let\approxgt=\gtrsim

\def\multleft#1{\hbox to size{\vbox {\halign {\lft{##}\cr #1}}\hfill}\par}
\def\multright#1{\hbox to size{\vbox {\halign {\rt{##}\cr #1}}\hfill}\par}

\def\boxit#1{\vbox{\hrule\hbox{\vrule\kern3pt\vbox{\kern3pt
          #1 \kern3pt}\kern3pt\vrule}\hrule}}

\def\erg{{\rm\thinspace erg}}

\def\Msun{\hbox{$\rm\thinspace M_{\odot}$}}

\def\s{{\rm\thinspace s}}

\def\ergps{\hbox{$\erg\s^{-1}\,$}}



\begin{document}

\title{Trapping of magnetic flux by the plunge region of a black hole
accretion disk}

\author{Christopher~S.~Reynolds\altaffilmark{1,2}, 
David~Garofalo\altaffilmark{3} and
Mitchell~C.~Begelman\altaffilmark{4,5,2}} 

\altaffiltext{1}{Dept.of Astronomy, University of Maryland, College
Park, MD20742} 
\altaffiltext{2}{KITP, University of California Santa Barbara, Santa Barbara CA93106}
\altaffiltext{3}{Dept.of Physics, University of Maryland, College
Park, MD20742} 
\altaffiltext{4}{JILA, University of Colorado, Boulder,
CO80309} 
\altaffiltext{5}{Dept. of Astrophysical and Planetary
Sciences, University of Colorado, Boulder, CO80309}

\begin{abstract}
The existence of the radius of marginal stability means that accretion
flows around black holes invariably undergo a transition from a MHD
turbulent disk-like flow to an inward plunging flow.  We argue that
the plunging inflow can greatly enhance the trapping of large scale
magnetic field on the black hole, and therefore may increase the
importance of the Blandford-Znajek (BZ) effect relative to previous
estimates that ignore the plunge region.  We support this hypothesis
by constructing and analyzing a toy-model of the dragging and trapping
of a large scale field by a black hole disk, revealing a strong
dependence of this effect on the effective magnetic Prandtl number of
the MHD turbulent disk.  Furthermore, we show that the enhancement of
the BZ effect depends on the geometric thickness of the accretion
disk.  This may be, at least in part, the physical underpinnings of
the empirical relation between the inferred geometric thickness of a
black hole disk and the presence of a radio jet.
\end{abstract}


\section{Introduction}

One of the most spectacular phenomena associated with accretion onto
black holes is the creation of powerful, highly-relativistic jets.
However, despite intense observational and theoretical study, the
basic energy source of these relativistic jets remains unknown.
Broadly speaking, there are two possibilities.  Firstly, jets could be
powered by the liberation of gravitational potential energy of
accreting matter.  If this is the case, the most likely scenario is
the formation and subsequent focusing and acceleration of a
magnetohydrodynamic (MHD) disk wind (Blandford \& Payne 1982).  While
this mechanism has the appealing feature of potentially being
universal to all accreting systems (and therefore allowing a unified
model for jets from proto-stellar systems, accreting white dwarfs and
accreting neutron stars as well as accreting black holes), it is not
clear that such a disk wind can be accelerated to the highly
relativistic velocities seen from many black hole systems.  The
alternative is that jets could be powered by the magnetic extraction
of the spin energy of the central black hole using the mechanism
described in the seminal paper by Blandford \& Znajek (1977; hereafter
BZ).  The power extracted from a Kerr black hole with dimensionless
spin parameter $a_*$ threaded by a magnetic field of strength $B_{\rm
H}$ (in the membrane paradigm sense; see Thorne, Price \& Macdonald
1986) is
\begin{equation}
L_{\rm BZ}\approx \frac{1}{32}\omega_{\rm F}^2B_{\rm H}^2 r_{\rm H}^2 a_*^2 c
\label{eqn:bz}
\end{equation}
where $r_H$ is the radius of the event horizon and
$\omega_F^2=\Omega_F(\Omega_H-\Omega_F)/\Omega_H^2$, with $\Omega_H$
and $\Omega_F$ being the angular velocities of the black hole and
magnetic field lines, respectively.  It is often argued (e.g., see BZ)
that the magnetic field structure adjusts itself such that
$\Omega_F=\Omega_H/2$ (Phinney 1983), hence maximizing $\omega_F^2$ to
a value of $1/4$.  While the initial work of BZ was based on
force-free black hole magnetospheres, the basic mechanism is seen to
operate in the recent generation of fully relativistic MHD accretion
disk simulations (e.g., see Koide et al. 2000; Komissarov 2004,
De~Villiers et al. 2004, McKinney \& Gammie 2004, McKinney 2005a,b,c)

In the past dozen years or so, several studies have cast doubt on
whether nature can produce significant hole-threading magnetic fields,
leading to the suggestion that the BZ mechanism is insufficient to
energize powerful black hole jets.  Developing upon work done by van
Ballegooijen (1989), Lubow, Papaloizou \& Pringle (1994) and Heyvaerts,
Priest \& Bardou (1996; hereafter HPB) have examined the dragging and
concentration of an external field by an MHD turbulent accretion disk.
Both sets of authors find that, due to the high effective magnetic
diffusivity of such disks, the inward dragging and subsequent
concentration of an external field is rather ineffective.  In a
different approach to this problem, Ghosh \& Abramowicz (1997;
hereafter GA97) construct force-free black hole magnetosphere models
within the radius of marginal stability and show that the field
threading the black hole is only of comparable strength to that
threading the inner disk.  Since the disk-threading field has to be
rather weak (with the magnetic pressure at least an order of magnitude
less than the total pressure) so as not to quench the
magneto-rotational instability (MRI) that drives the accretion itself
(Balbus \& Hawley 1991, 1998), the inferred black hole threading field
would be insufficient to energize the powerful jets in active galactic
nuclei (AGN) that we observe.  This argument was further developed by
Livio, Ogilvie \& Pringle (1999) who pointed out that, under these
circumstances, the electromagnetic power extracted from the inner
regions of the disk would necessarily dominate the black hole
spin-energy extraction.

In all of the studies described above, the plunge region of the black
hole accretion disk has been neglected.  This is the region of the
disk within the radius of marginal stability in which the accretion
flow is undergoing rapid inwards acceleration (ultimately crossing the
event horizon at the velocity of light as seen by a locally
non-rotating observer).  Unless the magnetic field is extremely
strong, this is a region where inertial forces will dominate and the
commonly employed force-free approximation will break down.  For
example, examination of the GA97 steady-state magnetosphere solution
shows magnetic field crossing the plunge region with a strength very
similar to that in the diffusive region of the disk.  This situation
seems unlikely --- any dynamically unimportant magnetic field that
threads the plunge-region would be swept into the black hole on a
dynamical timescale.  This means that the actual field threading the
plunge region would be very weak.  However, it does not imply that the
field threading the BH horizon, which is what counts for the BZ
effect, is also weak.  The field swept in by the plunge region would
be ``cleaned'' into some well-ordered configuration threading the
black hole (for a full discussion of the ``cleaning'' of a magnetic
field by a black hole, see Thorne, Price \& McDonald 1986) and can be
confined by the inertial forces of the plunging accretion flow even if
it achieves strengths appreciably higher than the characteristic field
strengths in the inner disk.  Since the strength of the BZ mechanism
depends on the square of the magnetic field, this enhancement could
have major implications for the relative dominance of spin-energy
extraction.

In this paper, we extend these previous works by examining the role of
the plunge region of a black hole accretion disk in enhancing the
horizon-threading flux.  In essence, we use the HPB formalism for flux
dragging in an accretion disk and impose an inner boundary condition
appropriate for the plunge region around a central black hole.  This
formalism is described in Section~2.  Although this analysis is
non-relativistic, it should provide guidance about flux enhancement by
the plunge region at least in the case of slowly rotating black holes.
As reported in Section 3, our analysis confirms the basic intuition
discussed in the previous paragraph and uncovers a strong dependence
of the equilibrium trapped flux on the disk thickness and the
effective magnetic Prandtl number of the disk.  Section~4 discusses
the sensitivity of our results to the outer boundary condition and
then places our results into a wider astrophysical context.  Our
conclusions are presented in Section 5.

\section{The toy model}

We follow the approach of HPB to study the dragging of an external
magnetic field by an MHD turbulent accretion disk.  As already noted,
this results in a non-relativistic model but should be able to provide
quantitative insights on the behavior of slowly rotating black holes
(where the radius of marginal stability is in a rather low
gravitational redshift regime).  We consider a thin Keplerian
accretion disk with geometric thickness $h\ll r$ extending down to the
radius of marginal stability at $r=r_{\rm ms}$.  We suppose that the
disk has an effective magnetic diffusivity (due to reconnection in the
MHD turbulence) of $\eta_*$ which is comparable to the effective
viscosity $\nu_*$.  In other words, the effective magnetic Prandtl
number $P_{\rm m}=\nu_*/\eta_*$ is of order unity (see HPB for an
explicit justification of $P_{\rm m}\sim 1$).  Note that we employ the
standard definition of ${P_{\rm m}}$ as the ratio of the viscosity to
the magnetic diffusivity which is the reciprocal of that used in Lubow
et al. (1994) and HPB.

We now suppose that an external uniform magnetic field with strength
$B_0$ is present in the vertical direction (i.e., aligned with the
normal to the accretion disk).  We are interested in the dragging of
this field by the accretion flow.  Assuming the system remains
axisymmetric at all times, and employing cylindrical polar coordinates
$(r,z,\phi)$, the poloidal magnetic field structure is completely
described by the flux function $A(r,z;t)$ via ${\bf B}_p=\nabla\times
(A\hat{\bf \phi}/r)$.  With such a definition, the magnetic flux
threading a ring of radius $r$ at height $z$ from the disk plane is
$2\pi A(r,z;t)$ and the magnetic field components are given by
\begin{eqnarray}
{\rm B}_r=-\frac{1}{r}\frac{\partial A}{\partial z},\\
{\rm B}_{\rm z}=\frac{1}{r}\frac{\partial A}{\partial r}.
\end{eqnarray}
The flux function can be decomposed into three components,
\begin{equation}
A(r,z;t)=A_{\rm BH}(r,z;t)+a(r,z;t)+\frac{r^2B_0}{2}
\label{eqn:adecomp}
\end{equation}
where $A_{\rm BH}(r,z;t)$ is the flux function associated with cleaned
black hole-threading field (generated by currents in the disk), the
final term on the RHS is just the uniform imposed flux (generated by
currents at infinity), and $a(r,z;t)$ accounts for all other
(disk-threading) magnetic field structures (generated, in principle,
by currents either in or out of the disk plane).  For definiteness we
suppose that, in the region exterior to the disk ($|z|>h$), the black
hole-threading field has the form of a split monopole,
\begin{equation}
A_{\rm BH}(r,z;t)=A_*(t)\left(1-{\rm
sgn}(z)\frac{z}{(z^2+r^2)^{1/2}}\right)\hspace{2cm}(|z|>h),
\label{eqn:splitmono}
\end{equation}
where $A_*(t)$ is $1/2\pi$ times the total hole-threading flux.  To
reiterate, this hole-threading field is generated by toroidal currents
flowing in the disk ($|z|<h$) and is a vacuum solution to Maxwell's
equations elsewhere.  While the precise structure of the cleaned black
hole field is unclear, the choice of the split monopole has support
from recent General Relativistic MHD simulations (e.g., see Hirose et
al., 2004; Komissarov 2005).

HPB showed that the time evolution of the flux function in the
diffusive part of the disk plane, $A(r,0;t)$ ($r>r_{\rm ms}$), is
given by
\begin{equation}
\frac{\partial A}{\partial t}+v_r\frac{\partial A}{\partial r}-\eta_*\left[\frac{1}{h}\left(\frac{\partial A}{\partial z}\right)_{\rm z=h}+r\frac{\partial}{\partial r}\left(\frac{1}{r}\frac{\partial A}{\partial r}\right) \right]=0,
\label{eqn:init_evol}
\end{equation}
where $v_r$ is the radial velocity of the accretion flow.  The
$(\partial A/\partial z)_{z=h}$ term represents the effect of magnetic
tension due to the curvature of field lines across the disk plane, and
thus depends on the structure of the magnetic field above and below
the disk.  For example, one could formulate steady-state MHD wind
solutions which take the instantaneous value of $A(r,0;t)$ as a
boundary condition.  This would be a task of great complexity (note
that the radial structure of the boundary condition would not admit
self-similar wind solutions).  Here, we make the following
simplifications.  Firstly, we assume that the magnetic field outside
of the disk (hereafter referred to as the disk magnetosphere) is
force-free, i.e., $(\nabla\times {\bf B})\times {\bf B}=0$.  Secondly,
we assume that the Alfv\'en speed in the disk magnetosphere is
sufficiently high as to reduce the toroidal field to essentially zero
(through the production of torsional Alfv\'en waves).  Setting
$B_\phi=0$, the field in the disk magnetosphere becomes potential
($\nabla\times {\bf B}=0$) and the flux function obeys
\begin{equation}
{\cal D}A=0
\end{equation}
where ${\cal D}$ is the linear differential operator
\begin{equation}
{\cal D}\equiv \frac{\partial}{\partial
r}\left(\frac{1}{r}\frac{\partial }{\partial r}\right)+
\frac{\partial}{\partial z}\left(\frac{1}{r}\frac{\partial}{\partial
z}\right).
\label{eqn:firstevol}
\end{equation}
Noting that both the imposed uniform field and (exterior to the disk)
the black hole-threading field $A_{\rm BH}(r,z;t)$ individually obey
${\cal D}A=0$, the structure of the disk magnetosphere is determined
by solving the potential problem for $a(r,z;t)$, i.e., ${\cal D}a=0$.

At this point, a brief discussion of our $B_{\phi}=0$ assumption
(which leads to the potential field condition) is in order.  In the
non-relativistic treatment here, we can always assume that the Alfv\'en
speed in the disk magnetosphere is large enough such that any twist in
the magnetic field is removed via a torsional Alfv\'en wave.  However, our
intent is to produce a toy-model for accretion onto a black hole so we
should be wary of an assumption that so explicitly relies on
non-relativistic physics.  In a fully relativistic treatment, the
force-free magnetosphere around a black hole accretion disk would be
described by the Grad-Shafranov equation (BZ; MacDonald \& Thorne
1982; Uzdensky 2004, 2005).  It is found that the poloidal field
structure depends on both the poloidal current distribution (which
gives rise to toroidal fields) and the field line rotation (due to the
fact that the field lines are frozen into an orbiting accretion disk,
for example).  In particular, the field structure is affected by the
presence of an inner and outer light cylinder.  Ultimately, a
relativistic version of our model should study the flux dragging and
the magnetization of the black hole including these physical effects.
Here, we simply note that detailed studies of non-rotating (or
slowly-rotating) black hole magnetospheres have shown that the field
line rotation associated with a Keplerian accretion disk has only a
small effect on the poloidal field as compared with the equivalent
non-rotating configuration (MacDonald 1984; Uzdensky 2004).  In this
sense, Keplerian accretion disks are ``slow rotators'' (Uzdensky
2004).

For the rest of this paper, we explicitly consider the behavior of the
magnetic field in the upper half of the $z$-plane, $z>0$ --- we
suppose the system to be symmetric in the $z=0$ plane such that
$B_z(r,z)=B_z(r,-z)$ and $B_r(r,z)=-B_r(r,-z)$.  The tension term in
eqn.~\ref{eqn:init_evol} can be decomposed into
\begin{equation}
\left(\frac{\partial A}{\partial z}\right)_{z=h}=\left(\frac{\partial
A_{\rm BH}}{\partial z}\right)_{z=h}+\left(\frac{\partial a}{\partial
z}\right)_{z=h}.
\label{eqn:tension}
\end{equation}
The contribution from the hole-threading flux can be evaluated
directly from eqn.\ref{eqn:splitmono},
\begin{equation}
\left(\frac{\partial A_{\rm BH}}{\partial z}\right)_{z=h}\approx
-\frac{A_*}{r},
\end{equation}
where we have neglected a term which is smaller by a factor of $(h/r)^2$.  The
remaining contribution to eqn.~\ref{eqn:tension} follows from the
solution to the potential problem ${\cal D}a=0$ with boundary
conditions $a(r=0,z;t)=0$ and $a(r,0;t)$ specified.  As shown by HPB,
this gives
\begin{equation}
\left(\frac{\partial A}{\partial z}\right)_{z=h}={\cal P}\int_0^\infty dx\frac{[a(x,0;t)-a(r,0;t)]}{\pi(r-x)^2}-\frac{a(r,0;t)}{\pi r}
\end{equation}
where ``$\cal P$'' signifies the principal part of the integral.  We
can now write an explicit integro-differential equation for the time
evolution of $a(r,0;t)$ in the diffusive part of the disk ($r>r_{\rm
ms}$);
\begin{eqnarray}
\frac{\partial a}{\partial t}+\frac{\partial A_*}{\partial t}+v_rrB_0+\left(v_r+\frac{\eta_*}{r}\right)\frac{\partial a}{\partial r} = \nonumber\\ \eta_*[\frac{1}{h}{\cal P}\int_0^\infty dx\frac{[a(x,0;t)-a(r,0;t)]}{\pi(r-x)^2}- \frac{a(r,0;t)}{h\pi r}-\frac{A_*(t)}{hr} +\frac{\partial^2a}{\partial r^2}].
\label{eqn:evol}
\end{eqnarray}

As part of our model, we must specify $h(r)$, $v(r)$ and $\eta_*(r)$.
For definiteness, we define $h(r)$ by taking the ratio $h/r$ as a
fixed parameter of our model (in principle, one could substitute a
particular form for $h(r)$ resulting from a detailed disk model).  To
specify the radial velocity field, we follow Lubow et al. (1994) and
split our disk into two zones which we dub an ``active'' and a
``dead'' zone.  In the active zone ($r_{\rm ms}<r<r_{\rm dead}$), we
set $v_r=-\nu_*(1/r-1/r_{\rm dead})$ where $\nu_*=\alpha
h^2(GM/r^3)^{1/2}$ (HRB), and $\eta_*=\nu_*/{P_{\rm m}}$.  The
magnetic Prandtl number ${P_{\rm m}}$ is a fixed and constant
parameter of the active disk.  Note that we have introduced the usual
$\alpha$ of accretion disk theory (in contrast with HPB who implicitly
employ $\alpha\sim 1$).  In the dead zone ($r_{\rm dead}<r<r_{\rm
out}$), the diffusivity is still given by $\eta_*=\alpha
h^2(GM/r^3)^{1/2}/{P_{\rm m}}$, but the velocity is set to zero.  For
computational necessities, we impose an outer cutoff on the system at
$r=r_{\rm out}$.  We assume that the disk beyond $r_{\rm out}$ is a
perfect and static conductor.  Hence the total magnetic flux threading
a loop ($r=r_{\rm out},z=0$) is constant and has the value $\pi r_{\rm
out}^2B_0$.  The inclusion of the dead-zone makes the evolution of the
inner part of the system essentially independent of the position or
exact nature of the $r=r_{\rm out}$ boundary.  In particular, the dead
zone acts as a reservoir of magnetic flux that can feed the actively
accreting part of the disk --- only in the outermost parts of the
active disk does the conservation of magnetic flux lead to a
non-negligible magnetic pressure trying to ``suck'' magnetic flux out
of the active disk.  The physical nature of the dead zone will be
discussed in Section~4.

Finally, we must specify boundary conditions on $a(r,0;t)$.  The
implementation of the inner boundary condition must capture the fact
that the plunge region is extremely effective at sweeping in poloidal
magnetic field that crosses within $r=r_{\rm ms}$.  Consider a
poloidal magnetic field line which is dragged towards the plunge
region on the viscous timescale $t_{\rm visc}\approx (r_{\rm
ms}/h_{\rm ms})^2(r_{\rm ms}^3/GM)^{1/2}\alpha^{-1}$.  Once in the
plunge region, the radial velocity of the disk material rapidly
increases with no associated increase in the effective magnetic
diffusivity (indeed, to the extent that the plunge region becomes a
laminar rather than a turbulent flow, the effective magnetic
diffusivity may well plummet to very small values).  For the field
strengths under consideration here (i.e., with an energy density much
less that the kinetic energy density of the accretion flow) inward
advection of the field line on a dynamical timescale $t_{\rm
dyn}\approx (r_{\rm ms}^3/GM)^{1/2}$ will dominate all other
processes.  Since the characteristic evolution timescale of the system
is $t_{\rm visc}\gg t_{\rm dyn}$, flux conservation gives that the
vertical magnetic field in the $z=0$ plane in the plunge region
compared with that in the disk just outside is
\begin{equation}
\frac{B_z({\rm plunge})}{B_z({\rm disk})}\approx \frac{t_{\rm dyn}}{t_{\rm
visc}}\approx\alpha\left(\frac{h}{r}\right)^2\ll 1.
\end{equation}
To a good approximation, we can say that the magnetic flux locally
crossing the plunge region is zero.  Thus, the only magnetic flux
passing through a loop $(r<r_{\rm ms},z=0)$ is that which threads the
black hole, i.e., $A(r\le r_{\rm ms},0;t)=A_*(t)$.  To cancel the
contribution from the externally imposed uniform field in this region,
we must have
\begin{equation}
a(r,0;t)=-r^2B_0/2\hspace{2cm}(r<r_{\rm ms}).
\label{eqn:ain}
\end{equation}
Thus, the appropriate inner boundary condition for eqn~\ref{eqn:evol}
is $a(r=r_{\rm ms},0;t)=-r_{\rm ms}^2B_0/2$ and we must use
eqn.~\ref{eqn:ain} in the evaluation of the integral term of
eqn.~\ref{eqn:evol}.  The fact that $\partial a(r_{\rm
ms},0;t)/\partial t=0$ allows us to use eqn.~\ref{eqn:evol} to
evaluate the rate of change of black hole-threading flux,
\begin{equation}
\frac{\partial A_*}{\partial t}= \eta_*(r_{\rm ms})\left[\frac{1}{h}{\cal
P}\int_0^\infty dx\frac{[a(x,0;t)-a(r,0;t)]}{\pi(r-x)^2}-
\frac{a(r,0;t)}{h\pi r}-\frac{A_*(t)}{hr} +\frac{\partial^2a}{\partial
r^2}+B_0\right]_{r=r_{ms}},
\label{eqn:dabhdt}
\end{equation}
where we have used the continuity of $\partial a/\partial r$ across
$r=r_{\rm ms}$ to combine the third and fourth terms on the left hand
side of eqn.~\ref{eqn:evol}.  We can justify this assumption of
continuity as follows.  Suppose that this derivative was {\it
discontinuous} across $r=r_{\rm ms}$, resulting in a discontinuity in
the strength of the vertical magnetic field.  This would lead to a
large magnetic pressure gradient and a very rapid rearrangement of
material until continuity was achieved.  We do note, however, that we
expect a rather narrow transition zone just outside of $r=r_{\rm ms}$
where vertical magnetic field goes from zero to the value
characteristic of the disk.  We must spatially resolve this
transition in our numerical model.

For the outer boundary condition, we set $a(r_{\rm
out},z=0;t)=-A_*(t)$ for some $r_{\rm out}>r_{\rm dead}$.  This
amounts to bounding the entire system by a perfect and static
conductor in the disk plane ($z=0$) for all $r>r_{\rm out}$, as
discussed above.

With these assumptions, eqns.~\ref{eqn:evol} and \ref{eqn:dabhdt}
completely describe the evolution of $a(r,0;t)$ and $A_*(t)$ from
some initial state once we fix the magnetic Prandtl number ${P_{\rm
m}}$, the disk thickness $h/r$, the characteristic radii of the
problem ($r_{\rm ms}$, $r_{\rm dead}$, $r_{\rm out}$), the external
field strength $B_0$, and the viscosity parameter $\alpha$.  In fact
$\alpha$ and $B_0$ are trivial parameters of the model, affecting only
the scaling of the time coordinate and the absolute normalization of
$a$, respectively.  Furthermore, the inclusion of the dead-zone makes
the evolution of the inner disk/field essentially independent of the
location of the outer boundary $r=r_{\rm out}$. Hence, the non-trivial
parameters describing this system are ${P_{\rm m}}$, $h/r$, and
$r_{\rm dead}$.   For our initial condition, we take
\begin{eqnarray}
a(r,z=0,t=0)=\left\{
\begin{array}{ll}
-r^2B_0/2 & (r<r_{\rm ms})\\
-r_{\rm ms}^2B_0/2 & (r\ge r_{\rm ms})
\end{array}
\right.
\end{eqnarray}
This amounts to saying that the initial currents flowing in the disk
are only those required to cancel the imposed uniform field in the
plunge region.

\section{Solution method and results}

We solve eqn.~\ref{eqn:evol} numerically by discretizing it on a
logarithmic grid with 200 zones from $r_{\rm ms}=6$ to $r_{\rm
out}=150$ with the dead-zone starting at $r_{\rm dead}=100$.  Here and
for the rest of this paper, radii will be given in units of
gravitational radii $GM/c^2$.  We treat the advective ($\partial
a/\partial t$) terms using the second order van~Leer (1977) method.
All other terms (including the principal part integral) are also
differenced to second-order spatial accuracy.  The time evolution is
achieved through a simple first-order explicit scheme.  To ensure
numerical stability, we set the time-step to be $dt=(1/dt_{\rm
ad}^2+1/dt_{\rm diff}^2+1/dt_{\rm field}^2)^{-1/2}$, where the
advective, diffusive and field time-steps are given by $dt_{\rm
ad}=0.5\min[\Delta r/(v+\eta_*/r)]$, $dt_{\rm diff}=0.5\min[\Delta
r^2/\eta_*]$ and $dt_{\rm field}=0.5\min[h\Delta r/\pi \eta_*]$.

Figure~1 shows the time-evolution of $A_*$ for the case of ${P_{\rm
m}}=2$ and various choices of $h/r$ from 0.01 to 0.16.  In all cases,
the flux threading the black hole grows from zero and achieves some
positive steady state.  In all cases, the final equilibrium flux
threading the black hole exceeds $\pi r_{\rm ms}^2B_0$ (corresponding
to $A_*=18B_0$), thereby establishing the basic fact that the plunge
region can aid in the accumulation of significant magnetic flux
through the black hole.  For thicker disks, the increased inward
advection of the field (due to the increased radial inflow speed of
the accreting matter) coupled with the decreased effectiveness of
field diffusion leads to significant enhancements of the black
hole-threading flux above this baseline value.  The dependence of the
equilibrium value of $A_*$ on disk thickness and magnetic Prandtl
number is shown in Fig.~1b. For small ${P_{\rm m}}$, the enhancement
of the hole-threading flux above the canonical value of $A_*=18B_0$ is
very small.  However, for ${P_{\rm m}}$ of order unity or higher,
there is a strong $h/r$-dependent enhancement.

\begin{figure}
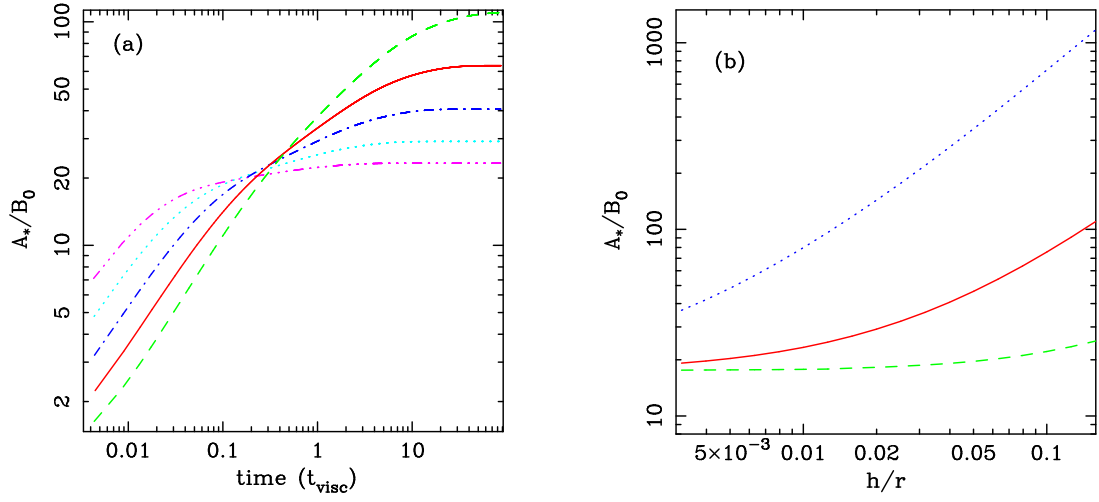

\hbox{
\hspace{0.5cm}
\psfig{figure=f1a.ps,width=0.4\textwidth,angle=270}
\hspace{1cm}
\psfig{figure=f1b.ps,width=0.4\textwidth,angle=270}
}
\caption{{\it Panel (a)} : Time dependence of the black hole-threading
flux for ${P_{\rm m}}=2$ and $h/r=0.01$ (magenta dot-dot-dot-dash line),
$0.02$ (cyan dotted line), $0.04$ (blue dot-dash line), $0.08$ (red
solid line), and $0.16$ (green dashed line).  For comparison,
$A_*/B_0=18$ corresponds to the flux of the uniform external field
threading the radius of marginal stability.  Time is in units of the
viscous timescale at $r_{\rm ms}$, $t_{\rm visc}=r^2(R^3/GM)/\alpha
h^2$.  {\it Panel (b)} : Equilibrium value of $A_*/B_0$ as a function
of $h/r$ for ${P_{\rm m}}=0.2$ (green dashed line), $2.0$ (red solid line)
and $20.0$ (blue dotted line).}
\end{figure}

\begin{figure}[t]
\hbox{
\psfig{figure=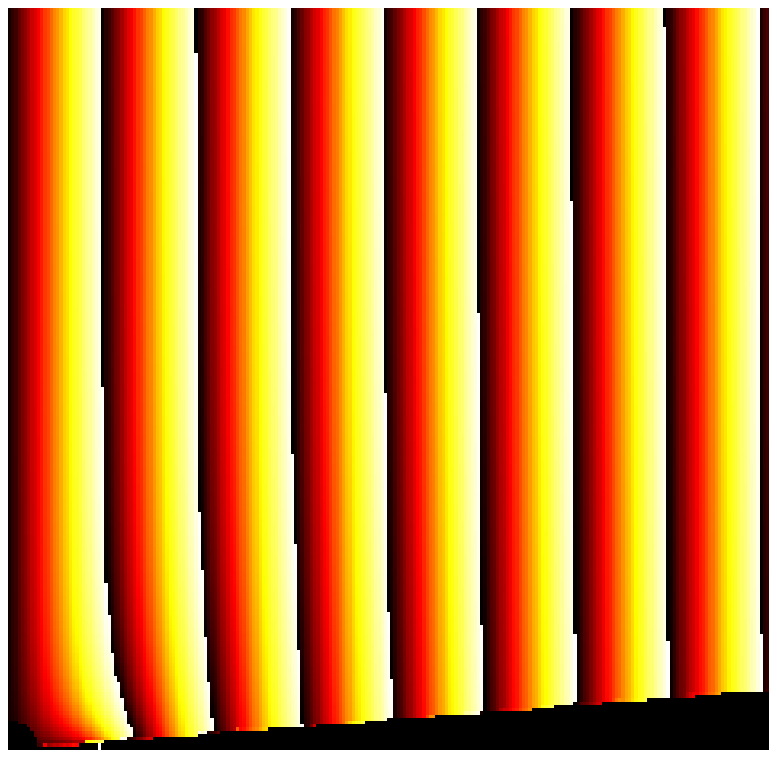,width=0.3\textwidth}
\hspace{0.25cm}
\psfig{figure=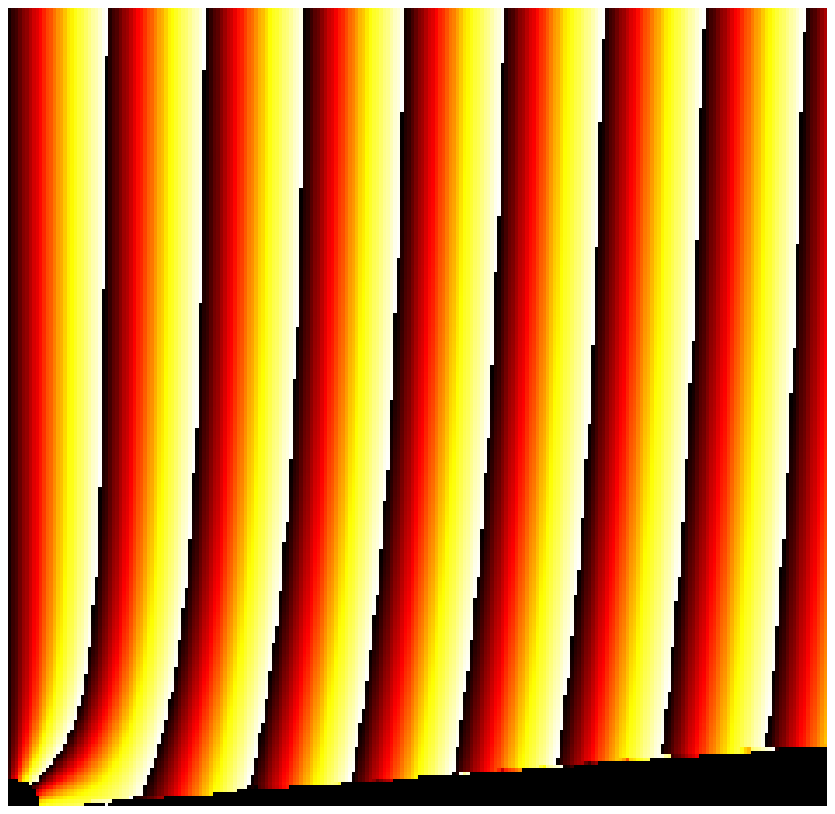,width=0.3\textwidth}
\hspace{0.25cm}
\psfig{figure=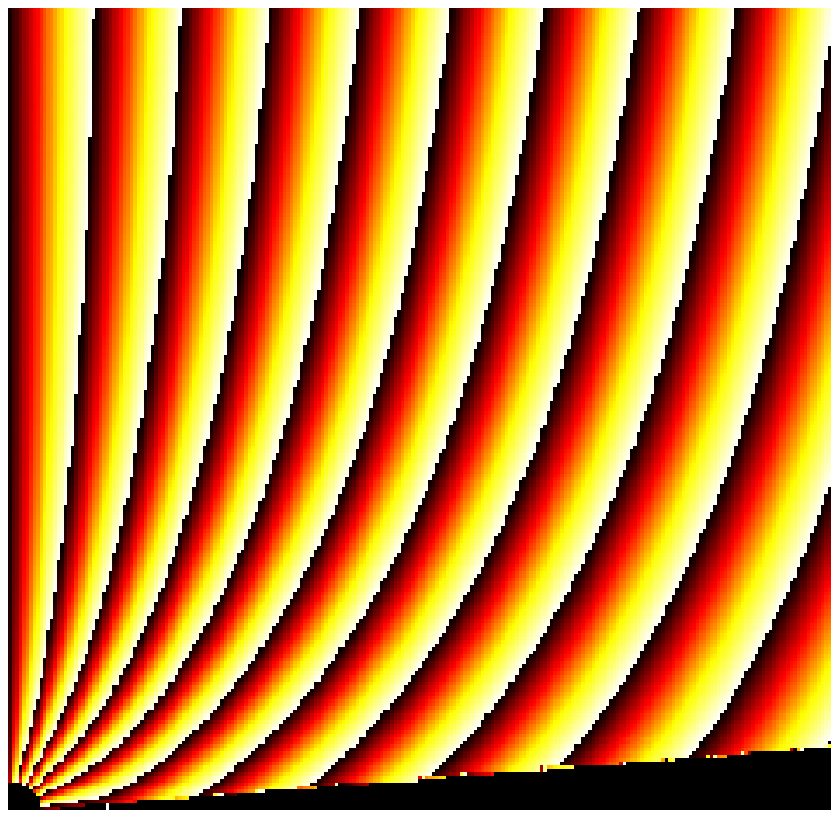,width=0.3\textwidth}
}
\caption{Magnetic field configuration for the initial condition (left
panel), the final state of the ${P_{\rm m}}=2$, $h/r=0.08$ case
(middle panel) and the final state of the ${P_{\rm m}}=20$, $h/r=0.08$
case (right panel).  Note how the higher magnetic Prandtl number
results in a powerful inward dragging of magnetic field and subsequent
magnetization of the black hole.  Each of these three panels is 50
gravitational radii ($50GM/c^2$) on a side.}  
\label{fig:field}
\end{figure}

The full magnetic field configuration can be derived by solving the
potential problem for $a(r,z;t)$ using the solution method laid out in
HPB.  In Fig.~\ref{fig:field}, we show the initial field configuration
as well as the final configuration for $h/r=0.08$ and two choices of
magnetic Prandtl number ${P_{\rm m}}=2$ and $20$.  The initial
configuration deviates from a simple uniform field due to the fact
that flux is excluded from the region $r<r_{\rm ms}$ which leads to a
``bowing'' of the field lines away from the radius of marginal
stability.  This curvature is rapidly reversed as field is advected
inwards, finally achieving a steady state in which the bend angle of
field lines as they enter the diffusive part of the disk is
approximately constant.  As pointed out by Lubow et al. (1994) and
discussed below, we expect this bend angle (away from the disk normal)
to be $i\sim\tan^{-1}(hP_{\rm m}/r)$.  This is indeed seen in our
equilibrium solutions.

The central quantity of interest in this work is the magnetic field
threading the black hole event horizon.  Recalling the definition of
the flux function, it is straightforward to show that the magnetic
field threading the event horizon is $B_{\rm H}=A_*/r_{\rm H}^2$ where
$r_{\rm H}=2$ is the event horizon radius of the (slowly rotating)
black hole considered in this work.  From the results described above,
we conclude that the equilibrium flux threading the black hole always
exceeds the flux of the external uniform field through the plunge
region ($\pi r_{\rm ms}^2B_0$ corresponding to $A_*=18B_0$), sometimes
by a large factor in the case of high effective magnetic Prandtl
numbers and/or thick disks.  Scaling to this fiducial flux, we have
$B_{\rm H}=4.5\Upsilon B_0$, where $\Upsilon=A_*/18 B_0$.  Using a
least squares fit to the results displayed in Fig.~1, we find that a
good approximation is $\Upsilon\approx 1+20{P_{\rm m}}(h/r)$.  Hence,
we have
\begin{equation}
B_{\rm H}\approx4.5\left[1+20{P_{\rm m}}\left(\frac{h}{r}\right)\right]B_0,
\label{eqn:bhfield}
\end{equation}
which is accurate to the 20\% level for ${P_{\rm m}}<20$.  As we
discuss below, the factor multiplying the $P_{\rm m}h/r$ term in
eqn.~\ref{eqn:bhfield} has a dependence on the size of the dead zone;
the precise form of eqn.~\ref{eqn:bhfield} is strictly valid only for
$r_{\rm dead}=100$.

\section{Discussion}

\subsection{Dependence on the size of the dead zone}

At first glance, the dragging of magnetic flux by the accretion disk
leads to a surprisingly large enhancement in the black hole-threading
field.  However, as we will now explain, simple arguments can be put
forward to support the results encapsulated in eqn.~\ref{eqn:bhfield}.

Firstly, we note that the existence of the dead zone is crucial for
setting an overall size scale to the magnetic disturbances introduced
by the disk.  To see this, consider the limit in which $r_{\rm
dead}\rightarrow \infty$ (also requiring $r_{\rm out}\rightarrow
\infty$, of course).  In this case, the imposed uniform magnetic
field is dragged inwards by the accretion flow but a balance will
never be achieved between the inward advection and the magnetic
tension --- without an imposed spatial scale, the field curvature
through the disk and hence the magnetic tension can be made
arbitrarily small.  A balance is possible only when one imposes an
outer truncation on the part of the disk that drags the magnetic flux.
In this case, the undragged field at $r>r_{\rm dead}$ acts as an anchor
and limits the vertical extent to which the magnetic field can be
appreciable distorted.  Indeed, our calculations show that the
magnetic field at $|z|>r_{\rm dead}$ is essentially just the imposed
uniform field.

Now, as already noted, we find that the magnetic field threads the
active part of the diffusive accretion disk ($r_{\rm ms}<r<r_{\rm
dead}$) with a bend angle (away from the disk normal) of $\tan i
\equiv B_r/B_z\approx hP_{\rm m}/r$.  As shown by HPB and Lubow et
al. (1994), this is a direct consequence of a balance between outward
magnetic diffusion due to field-line tension and the inwards advection
of magnetic field,
\begin{equation}
v_r\frac{\partial A}{\partial r}\approx\eta_*\left(\frac{\partial
A}{\partial z}\right)_{z=h}.
\end{equation}
Consider the field line that threads the inner edge of the diffusive
disk at $r=r_{\rm ms}$.  This field line follows a roughly parabolic
path in the magnetosphere that can be described by the flux function
$\Psi=\Psi_0(r^2+2\xi z)={\rm constant}$.  We can determine the
parameter $\xi$ using the fact that, at the disk plane, we have
$B_r/B_z\approx r/hP_{\rm m}$,
\begin{equation}
\frac{B_r}{B_z}=-\frac{\partial \Psi/\partial z}{\partial \Psi/\partial r}=-\frac{\xi}{r_{\rm ms}}\approx \left(\frac{h}{r}\right)P_{\rm m},
\end{equation}
where we have dropped a term that is second order in $(h/r)$.  At a
vertical distance of $z=r_{\rm dead}$, this same field line has a
cylindrical radius $R$ given by
\begin{equation}
R^2=r_{\rm ms}^2\left[1+2\frac{r_{\rm dead}}{r_{\rm
ms}}\left(\frac{h}{r}\right)P_{\rm m}\right]
\end{equation}
Using our observation above concerning the vertical extent of the
field disturbances, we use the fact that the field is essentially
uniform for $|z|>r_{\rm dead}$ to read off the magnetic flux threading
the plunge region and hence the black hole,
\begin{equation}
\Phi_{\rm H}=\pi R^2 B_0=\pi r_{\rm ms}^2 B_0\left[1 +2\frac{r_{\rm
dead}}{r_{\rm ms}}\left(\frac{h}{r}\right)P_{\rm m}\right].
\end{equation}
In terms of the field threading the hole (putting $r_{\rm ms}=6$) we get
\begin{equation}
B_{\rm H}=4.5\left[1 +\frac{r_{\rm
dead}}{3}\left(\frac{h}{r}\right)P_{\rm m}\right]B_0.
\end{equation}
Thus we can see that the numerical factor multiplying the $(h/r)P_{\rm
m}$ term in eqn.~\ref{eqn:bhfield} is directly related to the value of
$r_{\rm dead}$.

\begin{figure}
\hbox{
\psfig{figure=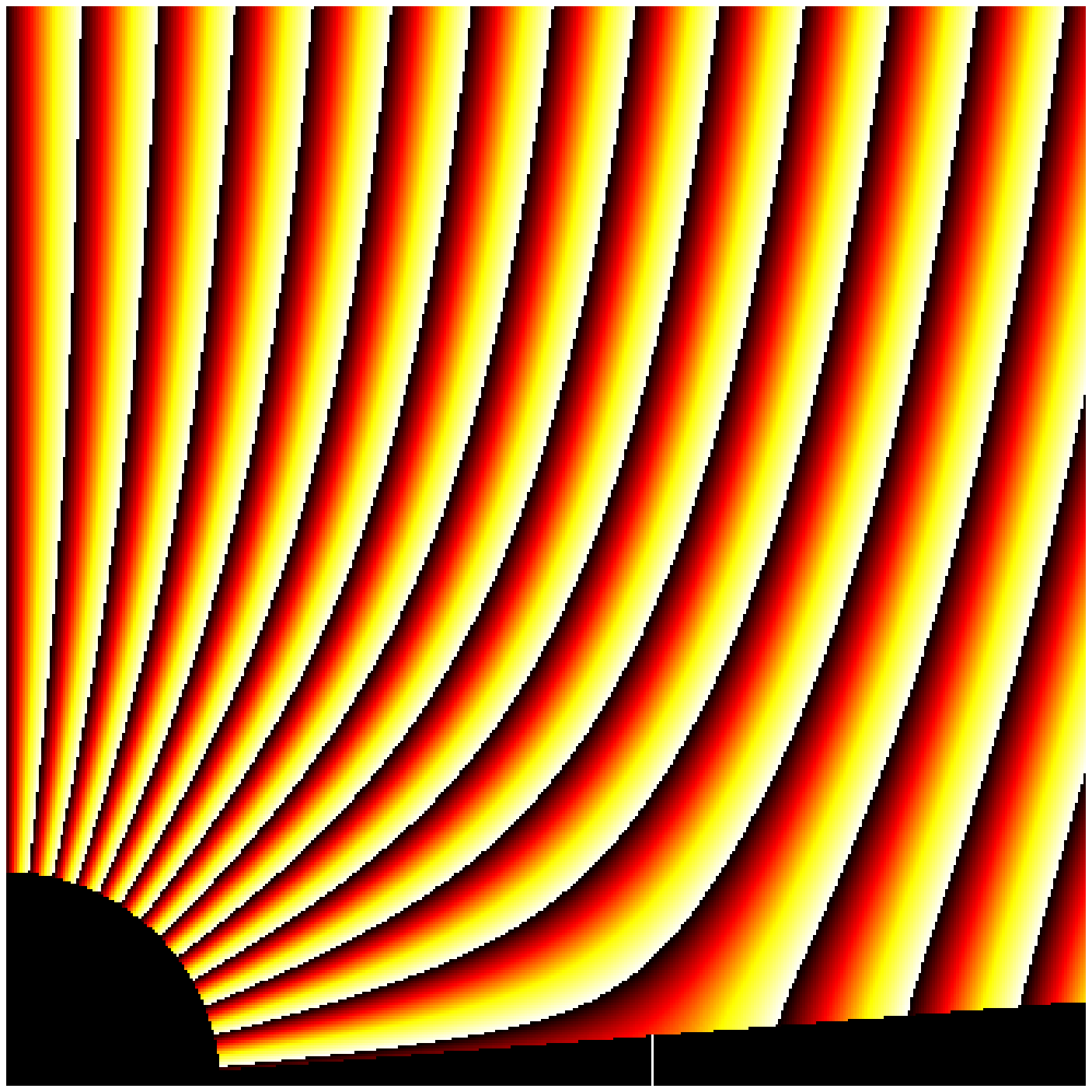,width=0.45\textwidth}
\hspace{0.5cm}
\psfig{figure=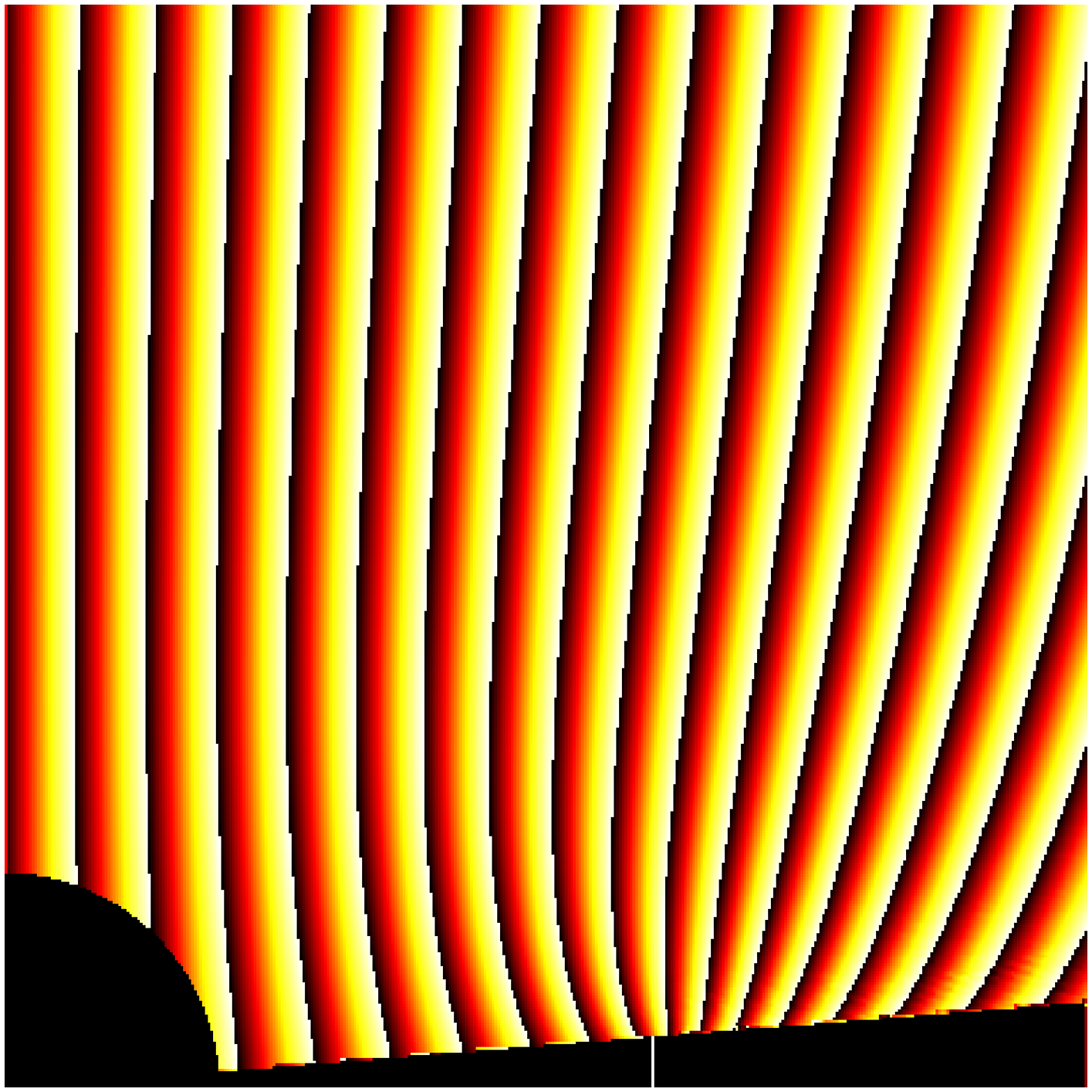,width=0.45\textwidth}
}
\caption{Magnetic field configurations for the plunge boundary
condition (left panel) and the uniform flux bundle boundary condition.
In both cases, the figure shows a zoom-in ($10GM/c^2$ on a side) of
the final field structure in the ${P_{\rm m}}=2$, $h/r=0.08$ case.  A
white vertical line on the accretion disk denotes the radius of
marginal stability.}
\end{figure}

The above discussion helps to elucidate the role of the plunge region in
enhancing the black hole-threading flux --- the plunge region
``shields'' the diffusive part of the disk from the large bundle of
magnetic flux that threads the black hole.  This bundle of flux is the
ultimate repository for the magnetic flux that has been scooped up the
by accretion flow.  The larger the region of the disk that can drag
the flux inwards, the larger is this repository.  To illustrate this
issue, we have run a modified version of our code in which the plunge
region boundary condition is replaced with the assumption that the
magnetic flux contained within $r=r_{\rm ms}$ has the form of a
uniform field on the disk plane.  We employ canonical values of the
model parameters; $h/r=0.08$, $P_{\rm m}=2$, and $r_{\rm dead}=100$.
As expected, we get a weak (50\%) enhancement in the flux contained
within $r=r_{\rm ms}$, compared with over a factor of 3 for the plunge
case.  The magnetic field structures of the two cases are illustrated
in Fig.~3.

Performing a full numerical solution to eqn.~\ref{eqn:evol} for
$r_{\rm dead}=50$, $r_{\rm dead}=100$ and $r_{\rm dead}=200$ reveals
that the enhancement of the magnetic flux increases with $r_{\rm
dead}$ slightly more slowly than the linear relationship predicted by
our simple arguments in this section.  Since the implementation of the
dead zone is one of most artificial aspects of our toy model, we will
not explore this dependence in any more detail in this paper.  In real
systems, the dead zone might be identified with the outer edge of the
MHD turbulence dominated accretion disk, e.g., the self-gravity region
in an AGN disk or the tidal truncation radius for the disk in a
Galactic Black Hole Binary (GBHB).  Both of these radii are likely to
be at significantly larger radius than $r_{\rm dead}=100$ used here.
Alternatively, if the magnetosphere is treated using a full MHD wind
model, the crucial length-scale which determines the magnetic field
enhancement is likely be the vertical height of the Alfv\'enic
surface.  It is beyond the scope of this paper to address such models.
However, our approach allows us to illustrate an essential point; that
the inward dragging of magnetic field over some region of the inner
disk coupled with the existence of the plunge region allows a
significant enhancement in the strength of the magnetic field
threading the black hole.

\subsection{Limitations of our approach}

Before discussing the astrophysical implications of our result, we
must address the three major limitations of our approach.  First, we
have made no attempt to include relativistic effects (beyond our
simple treatment of the radius of marginal stability) on the dynamics
or electrodynamics of the disk/field system.  Our model is an adequate
representation for slowly spinning black holes (where the radius of
marginal stability is rather large and in a comparatively low-gravity
region of spacetime), but we acknowledge that a full relativistic
electrodynamic treatment is required to robustly treat the case of
rapidly rotating black holes.  While the same basic phenomenon of
magnetic flux trapping by the plunge region should be at work around
rapidly rotating black holes, the geometry of the system (i.e., the
fact that the radius of marginal stability is much closer to the event
horizon) might be unfavorable for producing dramatic enhancements in
the black hole-threading magnetic field.  On the other hand, an
ergospheric wind (Punsly \& Coroniti 1990; Punsly 1991) could aid in
the production of a strong poloidal field (through the inertial
effects of the outflowing plasma) as well as the inward dragging of
field (through the removal of angular momentum from the accretion flow).

Second, we assume the existence of a pre-existing large scale magnetic
field.  The origin of such a field depends upon the system under
consideration.  For the accreting black hole at the heart of a
Gamma-Ray Burst (GRB) collapsar, such a field may arise naturally from
the collapsed stellar envelope.  In the case of AGN, the field
corresponds to that of the accreting interstellar medium.  For GBHBs,
the presence of a large scale field probably depends on the mode of
accretion, with wind-accretors likely possessing a much stronger and
better organized large scale field than Roche-lobe overflow accretors.

Third, we assume axisymmetric large scale fields with a disk
magnetosphere consisting of force-free and purely poloidal field.  As
mentioned in Section~2, a more physical treatment would entail
matching an MHD wind solution to the disk-plane flux function.  With
such an approach, one could capture the inertial effects of a disk
outflow on the field structure, the hoop stresses resulting from any
toroidal fields present, and the angular momentum losses in the disk
due to the wind.  These could have competing effects on the ultimate
ability of the disk to drag the field into the plunge region.  The
inertial effects will tend to bend the field lines outwards,
increasing the field-line curvature at the disk plane and hence
increasing outward diffusion of the field.  The loss of disk angular
momentum to the wind, on the other hand, would lead to an increase in
the radial velocity of the accretion flow but no change in the
magnetic diffusivity.  This, in turn, increases the inwards advection
of the magnetic field.  Clearly, more detailed calculations of this
scenario are warranted.  As for the axisymmetric assumption, we note
that Spruit \& Uzdensky (2005) have recently examined the dragging of
a large scale magnetic field by an accretion disk under the assumption
that the MHD turbulence in the disk concentrates the field into small
bundles (giving rise to the accretion disk equivalent of Sun spots).
Through an analysis of the dynamics of these bundles, they conclude
that this is a generally favorable scenario for accumulating a large
amount of magnetic flux in the central regions of the disk.  Thus, in
at least one specific model, an extreme deviation from axisymmetry
aids in the inward dragging of magnetic flux.

We reiterate that the principal result of this paper is that the
existence of a plunge region together with magnetic field dragging in
the accretion disk can significantly enhance the black hole-threading
magnetic field and hence the BZ effect.  Furthermore, the enhancement
becomes increasing effective for thicker disks or higher magnetic
Prandtl numbers.  However, we acknowledge that the caveats given above,
together with the dependence of the enhancement on the size of the
dead zone, prevents us from further quantifying the enhancement.

\subsection{Astrophysical implications}

Given the caveats discussed above, the results of Section~3 have
important implications for the strength of the black hole-threading
field and the relevance of the BZ process.  Suppose that the magnetic
pressure due to the large scale field $B_0$ is a fraction $f$ of the
maximum pressure in the accretion disk, $p_{\rm max}$, i.e.,
$B_0=(8\pi f p_{\rm max})^{1/2}$.  Using this together with
eqn.~\ref{eqn:bz} and eqn.~\ref{eqn:bhfield} gives, $L_{\rm BZ}\approx
5\pi\omega_F^2fp_{\rm max} \Upsilon^2r_{\rm H}^2a_*^2c$.  Using the
expressions for $p_{\rm max}$ for radiation pressure-dominated (RPD)
and gas pressure-dominated (GPD) disks from Moderski \& Sikora (1996)
and GA97, and assuming the usual BZ impedance matching criterion is
obeyed, gives
\begin{eqnarray}
L_{\rm BZ} (\ergps)\approx \left\{
\begin{array}{ll}
1.5\times 10^{45}\alpha^{-1}fM_8\Upsilon^2a_*^2 &{\rm RPD}\\
9\times 10^{43}\alpha^{-9/10}fM_8^{11/10}\dot{m}^{4/5}_{-4}\Upsilon^2a_*^2 &{\rm GPD}
\end{array}
\right.
\label{eqn:bzlum}
\end{eqnarray}
where we have scaled to a black hole mass of $M=10^8M_8\Msun$ and
$\dot{m}=10^{-4}\dot{m}_{-4}$ is the mass accretion rate in Eddington
units.  This can be directly compared with the expressions for $L_{\rm
BZ}$ in GA97 if we set $f\alpha^{-1}\approx 0.1$ (which results from
their relation between $\alpha$ and the disk magnetic field).  For
$\Upsilon=1$ (corresponding to small effective magnetic Prandtl
numbers or very thin disks), we find low BZ luminosities that agree
very well with those computed by GA97.  However, as we have shown,
large magnetic Prandtl numbers and/or thick disks can result in large
enhancements of the black hole-threading fields, approximately
described by $\Upsilon \approx 1+2xP_{\rm m}(h/r)$, where $x\sim {\cal
O}(r_{\rm dead}/r_{\rm ms})$.  The BZ luminosity is then enhanced by a
factor of $\Upsilon^2$.

It is interesting to explore astrophysical consequences of the strong
$h/r$ dependence of the equilibrium hole-threading flux $A_*$.  There
is mounting empirical evidence that black hole systems produce jets
only when a geometrically thick accretion disk is present.  The best
case can be made for the GBHBs, as discussed by Fender, Belloni \&
Gallo (2004).  In their X-ray low-hard (LH) state (a.k.a. the
power-law state; McClintock \& Remillard 2004) they display steady
optically-thick radio cores which, in Cygnus X-1, can be spatially
resolved into a jet-like structure by VLBA (Stirling et al. 2001).  It
is generally believed that the inner regions of the accretion flow in
a LH-state GBHB system is radiatively inefficient, hot, and hence
geometrically-thick ($h/r\sim 0.5$).  However, the radio jet is seen
to shut off once the source has made a transition to the high-soft
(HS) state (or thermal state; McClintock \& Remillard 2004) which is
believed to correspond to an inner accretion disk which is
radiatively efficient and hence significantly thinner.  We postulate
that the jet in the LH state is powered by the BZ effect which is
enhanced by the flux trapping effect of the plunge region.  Some time
after a transition to a HS state, the system will possess a disk with
a similar accretion rate but significantly reduced thickness.  For a
fixed accretion rate, the maximum pressure in a disk scales as $p_{\rm
max}\propto (h/r)^{-1}$.  Using our parameterization for $\Upsilon$,
we expect the BZ luminosity scales as $L_{\rm BZ}\propto f(h/r)$,
provided $h/r\approxgt 1/xP_{\rm m}$.  Hence, due to the inability of
a thin disk to trap flux on the black hole, the BZ luminosity of the
HS state will be much reduced leading to the suppression of the radio
jet.

The actual ${\rm LH}\rightarrow {\rm HS}$ transition itself is
particularly interesting.  It is during this transition (when the
source crosses the ``jet line'' on the X-ray flux/color diagram) that
powerful relativistic outflows are produced which, for example,
produce the superluminal radio blobs seen from microquasars.  It is
likely that the transition is driven by the thermal collapse of the
LH-state hot disk, producing a structure that eventually evolves into
the HS-state cold disk.  The nature of the intermediate structure is
unclear, however.  It has been suggested that the thermal collapse
produces a magnetically-dominated region (e.g., Meier 2005) in which
MRI-driven turbulence is suppressed and accretion proceeds only
through large scale magnetic torques.  If the pre-collapse disk is
threaded by a large scale magnetic field, this field could readily
become dynamically important in the post-collapse disk (since rapid
thermal collapse will proceed at constant surface density, producing a
disk pressure which scales as $p_{\rm max}\propto h/r$).  Subsequent
magnetic braking of the disk would lead to rapid inflow, a rapid
accretion of poloidal flux onto the black hole, and a rapid increase
in the importance of the BZ effect.  The powerful ejections seen from
GBHBs as they undergo this transition might be the result of such a
scenario.  The ejections would terminate once the inner disk has
ceased to be magnetically dominated (due to the accretion of matter
from the outer disk), hence re-establishing a turbulent state with
high effective magnetic diffusivity.

\section{Conclusions}

Black hole rotation is, in principle, a more than sufficient source of
energy for energizing even the most powerful relativistic jets.  The
viability of magnetic extraction of black hole spin energy does,
however, hinge on the strength of the horizon-threading poloidal
magnetic field that can be established by the accretion flow.  In this
paper, we have argued that the plunge region of the black hole
accretion disk has an important role to play in enhancing the
horizon-threading field well above the modest levels suggested by
previous works.  We support this hypothesis by constructing a
toy-model (that is non-relativistic, assumes axisymmetry, and treats
the fields away from the disk plane as potential) with which we can
follow the dragging of an external magnetic field by the disk and its
subsequent trapping by the plunge region.  Our toy model suggests that
the BZ effect can be enhanced above the canonical estimates of GA97 by
a factor of $[1+xP_{\rm m}(h/r)]^2$ where $P_{\rm m}$ is the effective
magnetic Prandtl number of the disk and $x\sim {\cal O}(r_{\rm
dead}/r_{\rm ms})$.  Even in cases where the effective magnetic
diffusivity is small due to the MHD turbulence (i.e., $P_{\rm m}\sim
1$), the BZ effect can be enhanced by one order of magnitude (or more)
above the GA97 value if the disk is geometrically-thick $h/r\approxgt
r_{\rm ms}/r_{\rm dead}$.  The $h/r$-dependence of this effect has an
appealing resonance with the empirical evidence from GBHBs which
points to a close connection between the existence of powerful black
hole jets and the inferred properties of the accretion disk.

\vspace{0.5cm}

We thank Phil Armitage and Cole Miller for insightful discussions
throughout this work.  This research was supported by the National
Science Foundation under grants PHY-990794 (CSR, MCB), AST-0205990,
and AST-0307502 (MCB).  Part of this work was carried out at the Kalvi
Institute for Theoretical Physics at the University of California,
Santa Barbara; CSR and MCB thank the members of KITP for their
hospitality.

\section*{References}

{\small

\noindent Balbus S.A., Hawley J.F., 1991, ApJ, 376, 214

\noindent Balbus S.A., Hawley J.F., 1998, Rev. Mod. Phys., 70, 1

\noindent Blandford R.D., Payne D.G., 1982, MNRAS, 199, 883

\noindent Blandford R.D., Znajek R.L., 1977, MNRAS, 179, 433 (BZ)

\noindent De~Villiers, J.P., Hawley J.F., Krolik J.H., Hirose S.,
2005, ApJ, 620, 878

\noindent Fender R., Belloni T., Gallo E., 2004, MNRAS, 355, 1105 

\noindent Ghosh P., Abramowicz M.A., 1997, MNRAS, 292, 887 (GA97)

\noindent Heyvaerts J., Priest E.R., Bardou A., 1996, ApJ, 473, 403 (HPB)

\noindent Hirose S., Krolik J.H., De Villiers J.P., Hawley J.F., 2004,
606, 1083

\noindent Koide S., Meier D.L., Shibata K., Kudoh T., 2000, ApJ, 536, 668

\noindent Komissarov S.S., 2004, MNRAS, 350, 1431

\noindent Komissarov S.S., 2005, MNRAS, 359, 801

\noindent Livio M., Ogilvie G.I., Pringle J.E., 1999, 512, 100

\noindent Lubow S.H., Papaloizou J.C.B., Pringle J.E., 1994, MNRAS,
267, 235

\noindent MacDonald D.A., 1984, MNRAS, 211, 313

\noindent MacDonald D.A., Thorne K.S., 198, 354

\noindent McClintock J.E., Remillard R.A., 2004, in 'Compact Stellar
X-ray Sources', eds. W.H.G.Lewin and M.~van~der~Klis, in press
(astro-ph/0306213)

\noindent McKinney J.C., 2005, ApJL, 630, L5

\noindent McKinney J.C., 2005, ApJ, submitted (astro-ph/0506368)

\noindent McKinney J.C., 2005, ApJ, submitted (astro-ph/0506369)

\noindent McKinney J.C., Gammie C.F., 2004, ApJ, 611, 977

\noindent Meier D., 2005, in 'X-ray Binaries to Quasars: Black Hole
Accretion on All Mass Scales', eds., T.J.Maccarone, R.P.Fender, L.C.Ho
(Dordrecht: Kluwer)

\noindent Phinney E.S., 1983, PhD thesis, Univ. of Cambridge

\noindent Punsly B., 1991, ApJ, 372, 424

\noindent Punsly B., Coroniti F.V., 1990, ApJ, 354, 583

\noindent Spruit H., Uzdensky D., 2005, ApJ, 629, 960

\noindent Stirling A.M., Spencer R.E., de la Force C.J., Garrett M.A.,
Fender R.P., Ogley R.N., 2001, MNRAS, 327, 1273

\noindent Thorne K.S., Price R.H., Macdonald D.A., 1986, Black Holes:
The Membrane Paradigm, Yale Univ. Press, New Haven CT

\noindent Uzdensky D., 2004, ApJ, 603, 652

\noindent Uzdensky D., 2005, ApJ, 620, 889

\noindent van Ballegooijen, A.A., 1989, in Belvedere G., ed.,
Accretion Disks and Magnetic Fields in Astrophysics, Kluwer,
Dordrecht, p.99

\noindent van Leer B., 1977, J. Comput. Phys., 23, 276

}

\end{document}